\documentclass[proof]{WileyASNA-v1}
\usepackage{geometry}
\usepackage{multicol,lipsum}
\usepackage{amsmath}
\usepackage{graphicx}
\usepackage{float}
\usepackage{wrapfig}
\usepackage{epstopdf}

\articletype{Original Paper}%

\received{<day> <Month>, <year>}
\revised{<day> <Month>, <year>}
\accepted{<day> <Month>, <year>}

\begin{document}

\title{Stellar and dark matter density in the Local Universe}

\author[1]{I.D. Karachentsev*}

\author[2]{K.N. Telikova}

\authormark{KARACHENTSEV \textsc{et al}}

\address[1]{\orgname{Special Astrophysical Observatory RAS}, \orgaddress{\state{Nizhnij Arkhyz, Karachai-Cherkessian Republic}, \country{Russia}}}

\address[2]{\orgname{Ioffe Institute}, \orgaddress{\state{26 Politekhnicheskaya st., St.\ Petersburg}, \country{Russia}}}

\corres{*I.D. Karachentsev, Special Astrophysical Observatory RAS, Nizhnij Arkhyz, Karachai-Cherkessian Republic, Russia. \email{ikar@sao.ru}}


\abstract{We calculate the mean density profiles for luminous and dark matter on distance scales $D \sim(1 - 100)$ Mpc around us using recent all-sky catalogs of galaxy groups. Within the Local Volume $( D < 11 ~\rm Mpc)$ we derived the mean stellar density $\Omega_*= 0.44\%$ in the critical density units and the mean total matter density $\Omega_m = 0.17$. In the sphere with a radius of 40 Mpc these quantities drop to $\Omega_* = 0.24 - 0.32\%$ and $\Omega_m = 0.09 - 0.14$. In a larger volume within $D\sim135$ Mpc the discussed densities become more uncertain: $\Omega_* = 0.20 - 0.24\%$ and $\Omega_m = 0.05 - 0.16$. We summarize that the major part of the cosmic dark matter locates outside the virial and collapsing zones of groups and clusters.}


\jnlcitation{\cname{%
\author{I.D. Karachentsev}, and 
\author{K.N. Telikova}} (\cyear{2018}), 
\ctitle{Stellar and dark matter density in the Local Universe}
}

 \DeclareGraphicsExtensions{.pdf}

\maketitle


\section{Introduction} 
Observational data on the structure and kinematics of the Local Universe is the widely used base for checking cosmological models. On a number of occasions this fact was pointed by \cite{Peebles1980,Peebles1993}. The paucity information on galaxy distances was a serious hindrance for the observational cosmology of the Local Universe for a long time. The situation has changed dramatically with the Hubble Space Telescope (HTS) commissioning. The unique abilities of the HST let astronomers separate the individual stars in the nearby galaxies and estimate the galaxy distances by the tip of the red giant branch (TRGB) method with $\sim(5-10)$\% accuracy. In the fast observational regime (one galaxy per unit orbit) it is available to measure distances for galaxies within 11 Mpc. At the present time the total number of galaxies with measured TRGB-distances in the Local Volume $(D<11~\rm Mpc)$ is about 400. These measurements are involved in the Updated Nearby Galaxy Catalog (UNGC) \citep{KarachentsevUNGC2013} and the Extragalactic Distance Database (EDD) \citep{Tully2009}.
 
 Outside the Local Volume the galaxy distances were estimated by Cepheid variables, type Ia supernovae and surface brightness fluctuations with $(5-10)\%$ accuracy (see EDD and references therein). Distances for about 5000 gas-rich galaxies were determined by Tully \& Fisher relation \citep{TullyFisher1977} between galaxy's luminosity and 21-cm emission line width with $\sim(20-25$)\% accuracy. Half of them locates within a distance of $\sim70$~Mpc.
 
\cite{Courtois2013} created maps of the large-scale distribution of galaxies in the Local Universe. These maps demonstrate the complicate density pattern produced by galaxy groups, clusters and empty areas. However, it is not easy to conclude from distribution of the attractors and voids surrounding the Milky Way whether our Galaxy is: in underdensity or overdensity region.

As far as we know, the first reconstruction of the mean density profile versus distance from the Milky Way was done by \cite{Makarov2011}. Authors calculated the stellar and total (virial) mass density up to $D\simeq45$ Mpc using a sample comprising 11000 galaxies with Galactic latitudes $|b|>15^{\circ}$. On these scales the estimated mean stellar density is greater than its global value. Nevertheless, the mean total density $\Omega_m$ in
the critical density units is systematically lower than cosmological value $\Omega_m = 0.24$ from WMAP \citep{Spergel2007} or $\Omega_m = 0.315$ from \cite{PlanckColl2014}.

The fact that the virial masses of nearby groups and clusters cannot provide the cosmological value of the matter density in the $\Lambda$CDM model has been already known.
According to independent estimates by \cite{Vennik1984} and \cite{Tully1987} the mean virial mass density inside the Local Supercluster is $\Omega_m\simeq0.08$, which $3-4$ times less than its global value.
Potential causes of this discrepancy were discussed in detail by \cite{Karachentsev2012}. Note, that recent papers by \cite{Tully2015a,Tully2015b,Kourkchi2017,Shaya2017} make an important contribution to ``the missing dark matter'' problem. 

In the next sections we present estimations of the mean luminous and total (dark) matter density on different scales from nearby widely investigated volume to farther poorly known regions in the Local Universe.

\section{Mean density profile in the Local Volume} 
The Updated Nearby Galaxy Catalog involves 869 galaxies with radial velocities $V_{LG}<600$~km~s$^{-1}$ or distances $D<11$~Mpc. Regularly updated online version of this database \citep{Kaisina2012} contains 1029 galaxies at the beginning of 2018 year\footnote{http://www.sao.ru/lv/lvgdb}.
  
Stellar masses in the UNGC were inferred from $K$-band luminosity of the galaxies as $M_* = (M_{\odot}/L_{\odot}) L_K$ \citep{Bell2003}. Majority of $K$-band magnitudes were measured in 2MASS Redshift Survey \citep{Jarrett2000}.
It is common knowledge that 2MASS misses
low surface brightness galaxies, especially with predominantly blue stellar population because of shot exposure time. For missing galaxies $K$-band magnitudes in UNGC were derived from $B$-magnitudes with respect to morphological types $T$ as \citep{Jarrett2003}

\begin{equation}
K=
 \begin{cases}
 	B-4.10, & \mbox{for } T<3 \\
    B-4.60+0.25 T, & \mbox{for } 3\leq T \leq 8 \\
    B-2.35, & \mbox{for } T>8 . 
 \end{cases}
\end{equation}
Two dozens high luminosity galaxies such as our Milky Way locate in the Local Volume. With respect to random orientation of the satellites' orbits and their mean eccentricity $\langle e^2\rangle =1/2$, the total mass of the parent galaxy halo can be defined as \citep{KarachentsevKudrya2014}

\begin{equation}
M_{tot}=(16/\pi) G^{-1} \langle\Delta V^2\ R_p\rangle, 
\end{equation} 
where $R_p$ is the projected separation between the dominated galaxy and its companion, $\Delta V$ is their radial velocity difference and $G$ is the gravitational constant. The sample of the luminous galaxies (Main Disturbers) with Galactic latitudes $|b|>15^{\circ}$ in the Local Volume is seen in Table~\ref{tab:MD}. Its columns contain: (1) galaxy name, (2) distance in Mpc, (3) radial velocity in the Local Group frame in km~s$^{-1}$, (4) logarithm of stellar mass in the solar mass units, (5) logarithm of halo mass, inferred from projected separation and radial velocity difference of the companions. Almost half of the galaxy population in the Local Volume belongs to satellites of these luminous galaxies.  
\begin{center}
\begin{table}[!ht]
\centering
\caption{\label{tab:MD} Luminous galaxies at $|b| > 15^\circ$ in the Local Volume.}
\begin{tabular}{lccll} 
\hline
Galaxy &$D$&$V_{LG}$&$\log M_*$&$\log M_{tot}$\\
&Mpc&km~s$^{-1}$&$M_{\odot}$&$M_{\odot}$\\
\hline
Milky Way&0.01&-65&10.70&12.07\\
M31&0.77&-29&10.79&12.23\\
NGC5128&3.68&310&10.89&12.89\\
M81&3.70&104&10.95&12.69\\
NGC253&3.70&276&10.98&12.18\\
NGC4826&4.41&365&10.49&10.78\\
NGC4736&4.41&352&10.56&12.43\\
NGC5236&4.90&307&10.86&12.02\\
M101&6.95&378&10.79&12.17\\
NGC4258&7.66&506&10.92&12.50\\
NGC3627&8.32&579&10.82&12.16\\
M51&8.40&538&10.97&11.78\\
NGC2903&8.87&443&10.82&11.68\\
NGC5055&9.04&562&11.00&12.49\\
NGC4594&9.55&894&11.30&13.45\\
NGC6744&9.51&706&10.91&11.72\\
NGC3115&9.68&439&10.95&12.54\\
NGC2683&9.82&334&10.81&12.13\\
NGC891&9.95&736&10.98&11.90\\
NGC628&10.2&827&10.60&11.66\\  
NGC3379&11.0&774&10.92&13.23\\
\hline
\end{tabular}
\end{table}
\end{center}       
      
\begin{figure*}[!ht]
\centerline{\includegraphics[width=342pt]{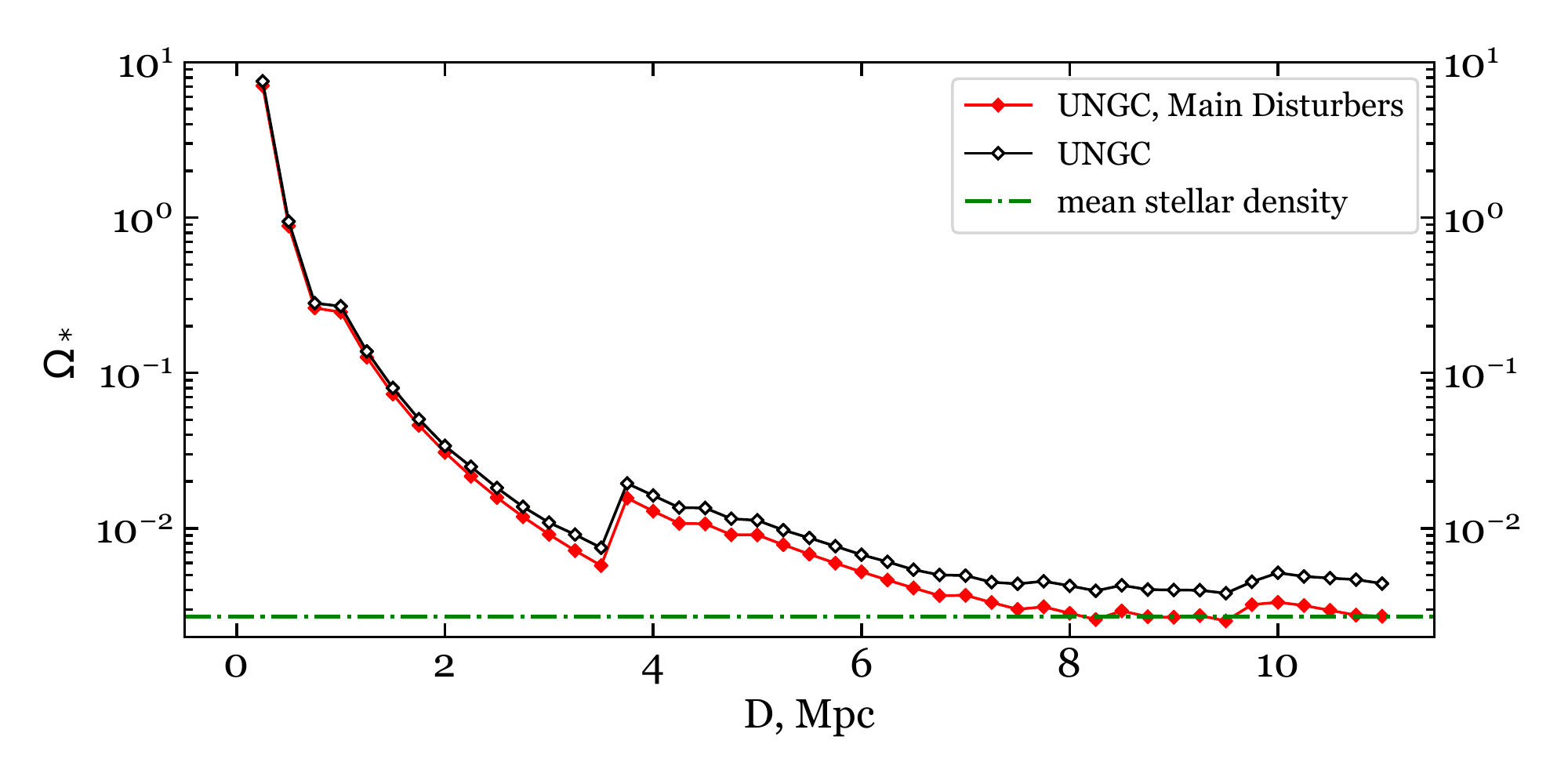}}
\caption{Mean density of stellar matter within a distance $D$ in the Local Volume. Empty diamonds show the stellar density of all galaxies in  the UNGC, filled diamonds correspond to the stellar density produced by 21 Main Disturbers seen in Table~\ref{tab:MD}. Dash-dotted horizontal line shows the global cosmic stellar density from \cite{Fukugita2004}. } 
\label{1}
\end{figure*} 
Figure~\ref{1} shows the mean stellar density in the Local Volume as a function of distance from the Milky Way. The global value of stellar density $\Omega_{*c}=0.0027\pm0.0005$ \citep{Fukugita2004} in the critical density units in Figure~\ref{1} is in a good agreement with the mean $K$-luminosity density $j_K=(4.3\pm0.2) \times 10^8~L_{\odot}$~Mpc$^{-3}$ from \cite{Jones2006,Driver2012}. The critical density can be expressed via the Hubble parameter $H_0$ as
\begin{equation}
\rho_C=\frac{3H^2_0}{8\pi G},
\end{equation}
consequently $\rho_C=1.0 \times 10^{-29}~\rm~g~cm^{-3}$ or $1.46\times10^{11}~M_{\odot}~\rm Mpc^{-3}$ for $H_0=73$~km~s$^{-1}$~Mpc$^{-1}$. Here and in the sections below we use a prefactor $(1-\sin15^{\circ})^{-1}\simeq1.35$ to compensate missed galaxies at $|b|<15^{\circ}$.
The total stellar mass of the 21 high luminosity galaxies is $1.6\times10^{12}~M_{\odot}$  or 59\% from the total stellar mass of the whole Local Volume sample. 
Notice, that stellar density on all the scales $D<11$~Mpc is greater than its global cosmic value. Peak at 3.7 Mpc is the
result of three nearby massive groups: NGC~5128, M~81 and NGC~253 with equally distances from the Milky Way (see Table~\ref{tab:MD}).

\begin{figure*}[!ht]
\centerline{\includegraphics [width=342pt]{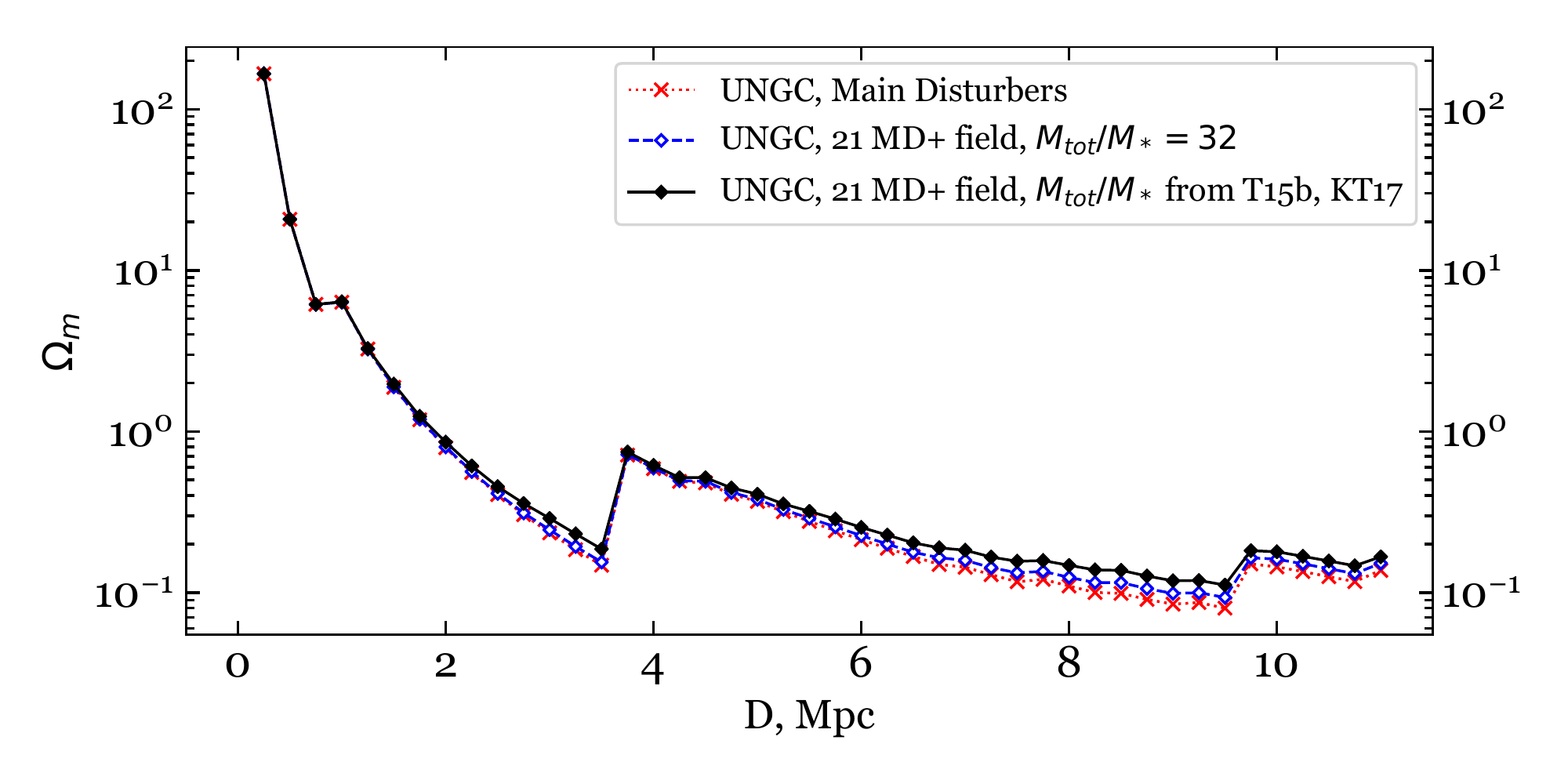}}
\caption{Mean density of dark matter within a distance $D$ in the Local Volume. Crosses show the dark matter density from 21 Main Disturbers. Filled and empty diamonds  show the dark matter density with account of all galaxies in the UNGC. Here the total masses of field galaxies were estimated under the assumption that $M_{tot}/M_*\simeq32$ \citep{KarachentsevKudrya2014} or using eq.~(\ref{M_L_Tully}) \citep{Tully2015b,Kourkchi2017} and shown by empty and filled diamonds, respectively.} 
\label{2}
\end{figure*}

Figure~\ref{2} shows the mean dark matter density in the Local Volume within a distance $D$ from the Milky Way. In the current cosmological scenarios the star formation process is the most efficient for stellar masses $M_*\simeq 1.0 \times 10^{9-10}~M_{\odot}$ \citep{Trujillo-Gomez2011}. This feature is accompanied with increase of the mass-to-light ratio, $M_{tot}/M_*$, towards luminous as well as faint galaxies. \cite{Kourkchi2017} provided an analytical approximation for mass-to-light ratio with such two branches :
\begin{equation}\label{M_L_Tully}
\log(M_{tot}/M_*) = 
 \begin{cases}
   \log(32)-0.50 \log(M_*/10^{10}), \\ \text{\hspace{1.5cm} for $\log M_*<8.97$}\\
   \log(32)+0.15 \log(M_*/10^{10}), \\ \text{\hspace{1.5cm} for $\log M_*>10.65$ .}
 \end{cases}
\end{equation}

Estimation of the dark matter density based on eq.~(\ref{M_L_Tully}) has a bit greater value than that based on $M_{tot}/M_*\simeq32$  relation following from Table~\ref{tab:MD} data. The total mass of the Local Volume turns out to be $10^{14}M_{\odot}$ with the respective mean dark matter density $\Omega_m$ = 0.17.

It can be seen from Figures~\ref{1} and \ref{2} that the profile of $\Omega_m(D)$ is similar to $\Omega_*(D)$. Naturally, this result is expected because of the key contribution of the 21 luminous galaxies to $\Omega_*$ and $\Omega_m$.

\section{Mean density profile in the Local Supercluster, $\mathbf{z<0.01}$} 
For 11 000 galaxies with radial velocities $V_{LG}<3500$~km~s$^{-1}$ at $|b|>15^{\circ}$ Makarov \& Karachentsev applied a new group-finding algorithm. In contrast to ``Friends of Friends'' percolation algorithm \citep{Huchra1982}, authors took into account a vast luminosity difference existing among galaxies \citep{Makarov2011}. They assumed that virtual galaxy pair has negative total energy and pair's members have a crossing time less than the age of the Universe. The varying parameter of the clusterization was calibrated with nearby galaxy groups. Using this procedure authors created catalogs of 509 galaxy pairs \citep{Karachentsev2008}, 168 triplets \citep{Makarov2009} and 395 galaxy groups with more then 3 members \citep{Makarov2011}. As a result, the catalogs contain 54\% of the initial galaxy sample or 82\% of its total $K$-band luminosity.
    
\begin{figure*}[t]
\centerline{\includegraphics [width=342pt]{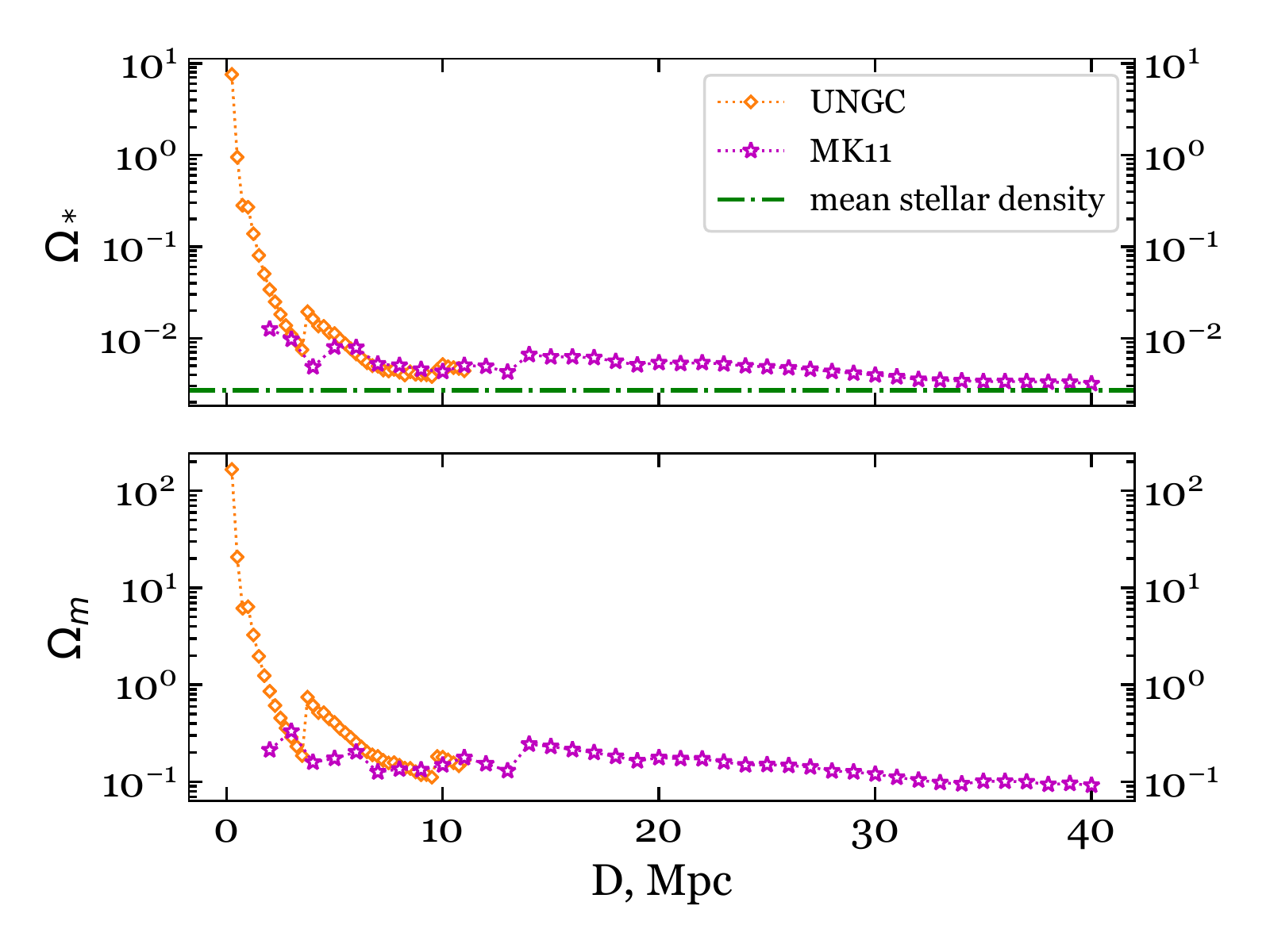}}
\caption{Mean density of the stellar matter (upper panel) and dark matter (lower panel) within a distance $D$ up to 40~Mpc. Diamonds and stars show the mean density from UNGC (Local Volume) and MK11 data, respectively. Dash-dotted horizontal line in the upper panel shows the global cosmic stellar density.} 
\label{3}
\end{figure*}

\begin{figure*}[t]
\centerline{\includegraphics [width=342pt]{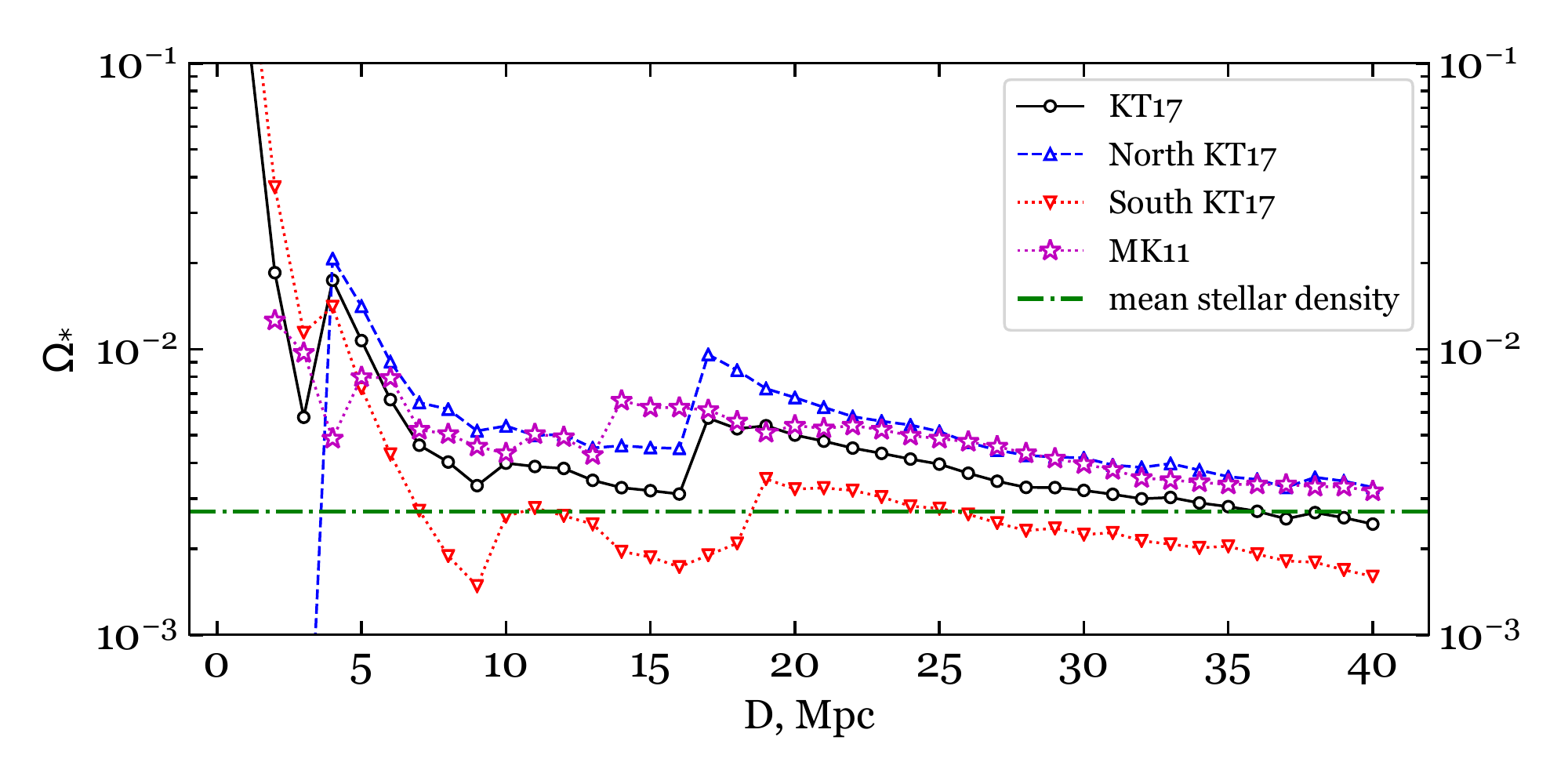}}
\caption{Stellar density within a sphere of radius $D$. Different symbols show the mean density based on KT17 catalog (circles for all sky, triangles for Northern and Southern Galactic hemispheres) and results by MK11 (stars). Dash-dotted horizontal line shows the global stellar density.} 
\label{4}
\end{figure*}        

Figure~\ref{3} shows profiles of the mean density of stellar matter (upper panel) and dark matter (lower panel) within different distances up to $D=40$~Mpc calculated in the Local Volume (Figure~\ref{1} and \ref{2}) and by \cite{Makarov2011} (hereinafter referred to as MK11). All density profiles in this paper are calibrated with the Hubble parameter  $H_0=73$ km~s$^{-1}$~Mpc$^{-1}$. The mean stellar density on the distance scales $D < 40$~Mpc is systematically greater than that of global cosmic value. Estimated total stellar mass and mean stellar density within this volume are $9.2\times 10^{13}M_{\odot}$ and 0.32\% in the critical density units, respectively. Taking into account that the galaxy distances in MK11 were estimated simply by their radial velocities, we conclude a good agreement between these two independent $\Omega_*$ sequences within the Local Volume. 
In contrast to $\Omega_*$, the mean density of the virial mass on the scales $D > 6$~Mpc is lower than its cosmological value. Within a sphere of radius $D=40$~Mpc, the total virial mass is $2.7\times10^{15}M_{\odot}$. The mean density on this scale decreases to $\Omega_m=0.09$. The secondary peak in the both panels at $\sim 14$~Mpc is the result of the Virgo cluster contribution with its virial mass of $6.3\times 10^{14}M_{\odot}$ \citep{Shaya2017}.

\cite{Kourkchi2017} (hereafter KT17) recently published a new catalog of galaxy groups with the same limits on radial velocity $V_{LG}< 3500$ km~s$^{-1}$ and Galactic latitude $|b|>15^{\circ}$ as in MK11. For clusterization algorithm KT17 used some empirical relations between the virial radius, velocity dispersion and the total mass of groups. Authors provided two types of estimations of the total group mass. Dynamic masses were inferred from radial velocity dispersion $\sigma^2_p$  and the mean harmonic radius of the group $R_h$:

\begin{equation}
M_{dyn}=(\alpha\pi/2G) \sigma^2_p R_h,
\end{equation} 
where parameter $\alpha=2.5$ is written to account for projection effects. Paucity of knowledge about the kinematics of distant galaxies implies significant uncertainties of $M_{dyn}$. That is why KT17 applied also another mass estimate. To determine the halo mass from a galaxy stellar mass (or $L_K$-luminosity) they used eq.~(\ref{M_L_Tully}). All scaling relations were calibrated with 8 nearby galaxy groups. In the clusterization criterion authors took into account the significant diversity of galaxies' luminosities. As a result, KT17 applied their algorithm to 15004 galaxies and created the catalog of 1536 galaxy groups which is presented in EDD\footnote{http://edd.ifa.hawaii.edu}. About 49\% of the total sample still remained as isolated galaxies. We used this catalog to investigate the stellar and virial mass distribution on the distance scales $D<40$~Mpc. For unification we restate data from KT17 for the same Hubble parameter $H_0=73$~km~s$^{-1}$~Mpc$^{-1}$.

Figure~\ref{4} presents the mean stellar density profile up to $D=40$~Mpc based on KT17 catalog. There is a systematic difference
between the mean density within Northern and Southern Galactic hemispheres. The secondary peak at 16.5 Mpc in the Northern cap is caused by the Virgo cluster, and the Fornax cluster gives a secondary peak on the Southern branch near $D = 19$ Mpc. In this Figure we also added the MK11 results. In almost all the bins $\Omega_*$ from KT17 catalog is slightly lower than the mean density in MK11. The ratio $\Omega_*^{KT17} / \Omega_*^{MK11}$ close to 0.77 both at the edge of the Local Volume and at $D=40$~Mpc. We suppose that MK11 infer the total flux from bluish diffuse galaxies, missed in 2MASS survey, more accurately. 
Note, that the Virgo cluster peak in the MK11 data is shifted from
16.5 Mpc to 14 Mpc. The kinematic distance of the Virgo cluster in the MK11 catalog has been determined via the mean cluster velocity, which is biased on $-200$~km~s$^{-1}$ due to falling the Local Group towards the Virgo. 
The variations between the different curves in the Figure~\ref{4} give a visual
representation of systematic effects affecting the estimation of the mean local density of matter.

\begin{figure*}[t]
\centerline{\includegraphics [width=342pt]{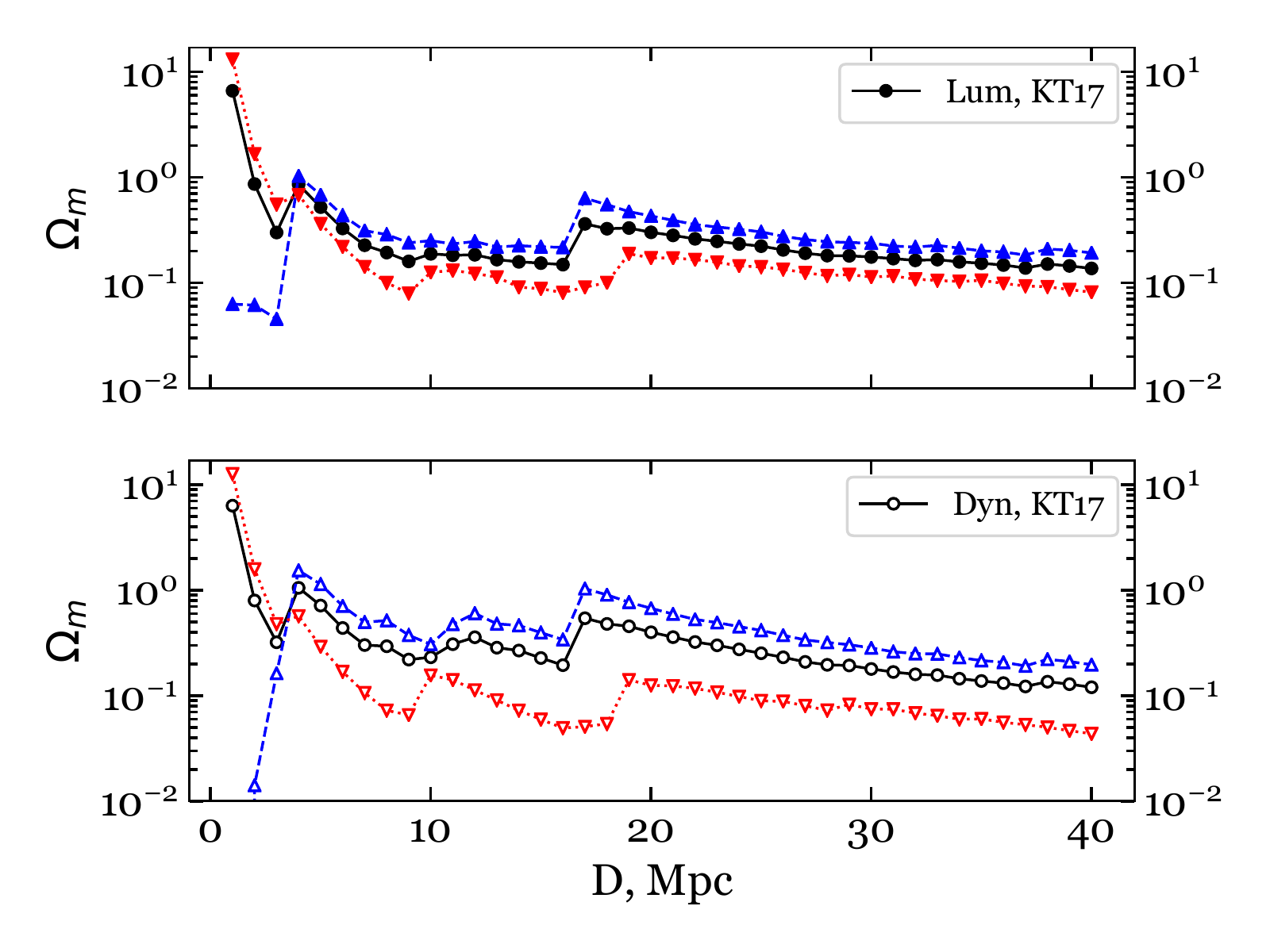}}
\caption{Mean density of the total mass, calculated by luminosity (upper panel) and by dynamic mass (lower panel) from KT17. Different symbols show the mean density for the whole sky (circles) and for Northern/Southern hemisphere (triangles).} 
\label{5}
\end{figure*}

Behaviour of the dark matter's mean density within a sphere of radius $D$ is shown in Figure~\ref{5}. Its upper and lower panels present the virial masses estimated by eq.~(\ref{M_L_Tully}), $M_{Lum}$, and by galaxy group's kinematic properties, $M_{dyn}$, respectively. Note a significant difference between the $\Omega_m$ for Northern and Southern Galactic hemispheres at all the distances $D<40$~Mpc. For mass inferred from galaxy group's luminosity this difference is less distinct than for dynamic mass. We conclude that the Local Universe within $\sim40$~Mpc still not fit the size of cosmic homogeneity cell. 

\begin{figure*}[t]
\centerline{\includegraphics [width=342pt]{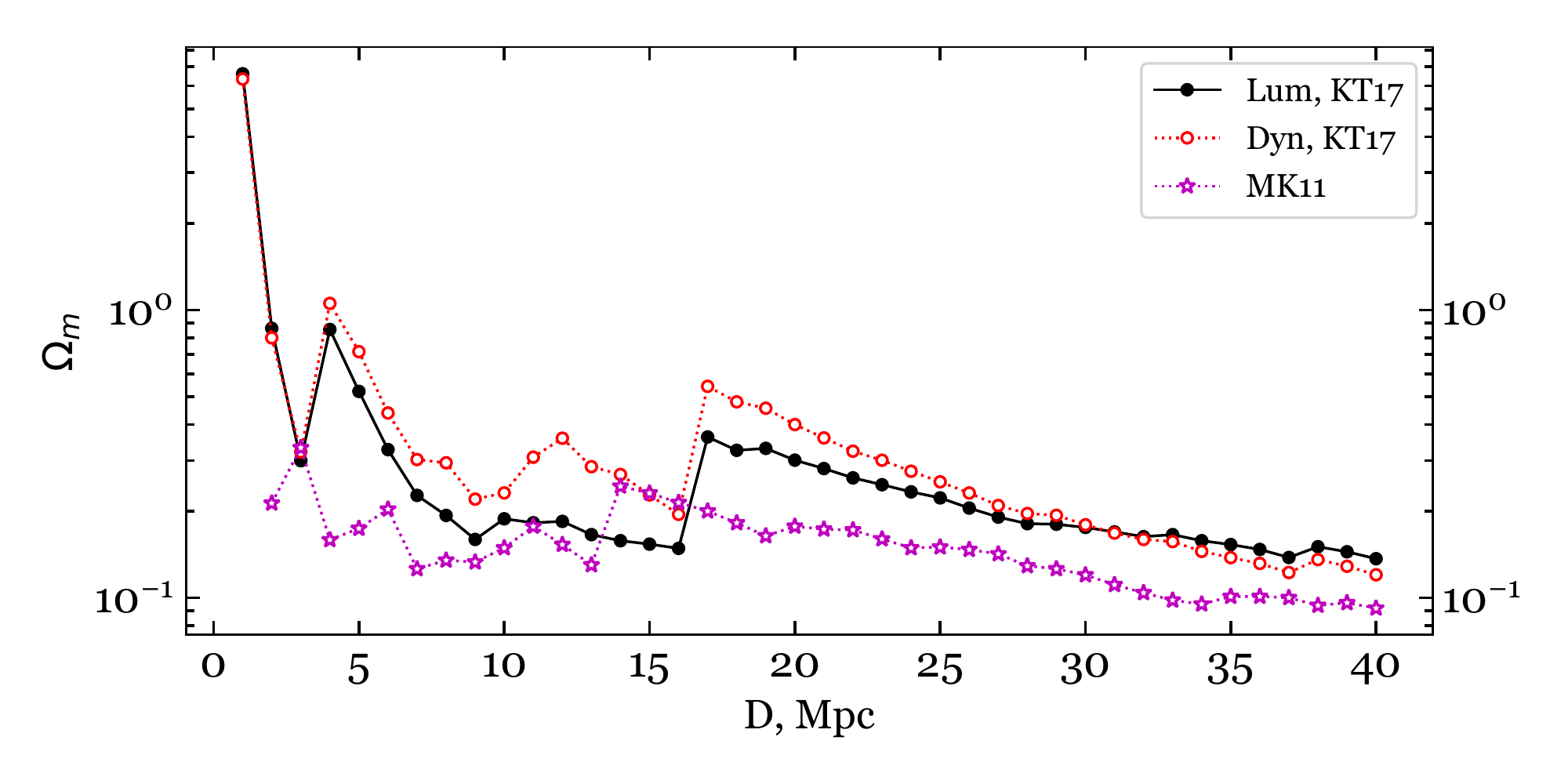}}
\caption{Mean density of the total mass inferred from luminosity of the galaxy groups (filled circles) and from dynamic masses (open circles). Data from MK11 catalog is shown by stars.} 
\label{6}
\end{figure*}

In Figure~\ref{6} we summarize three independent estimations of the virial mass' mean density within 40~Mpc. In MK11 catalog authors take into account also masses of triple, binary and isolated galaxies. At the $D<20$~Mpc scales these estimations of $\Omega_m$ differ from each other significantly, but at the edge of considered volume the mean densities lie in the narrow range $\Omega_m = 0.09-0.14$, showing a trend to further decreasing. The total mass within a sphere of radius $D=40$~Mpc is $(2.7-4.0)\times 10^{15}M_{\odot}$ with the Virgo cluster contribution as $(16-23)\%$.

\section{$\Omega_*$ and $\Omega_m$ within 10000~km~s$^{-1}$}  
Based on the 2MASS Redshift Survey \citep{Huchra2012} containing objects with magnitudes up to $K_s=11.75^m$, \cite{Tully2015b} (hereafter T15b) created a catalog of  galaxy groups with $V_{LG}=3000-10000$~km~s$^{-1}$. The clusterization algorithm of galaxies was the same as in the closer volume \citep{Tully2015a}. About 58\% of the total sample accounting 24044 galaxies were clustered into 3461 groups with two or more members.  
          
It is obvious that on long distances 2MASS Redshift Survey misses a significant number of galaxies because of the bright observational limit $K_s=11.75^m$. Taking this fact into account, Tully calculated a correction factor (CF) for the total luminosity of a group. To estimate the CF, Tully assumed that galaxy luminosity function is well described by Schechter function \citep{Schechter1976} with parameters $\alpha_K=-1.0$ and $M^*_K=-24.23$ for $H_0$ = 73~km~s$^{-1}$~Mpc$^{-1}$. Resulted CF-factor is negligible at $V_{LG}<1300$~km~s$^{-1}$, but increases to 2.3 at $V_{LG}=10000$~km~s$^{-1}$.

\begin{figure*}[t]
\centerline{\includegraphics [width=342pt]{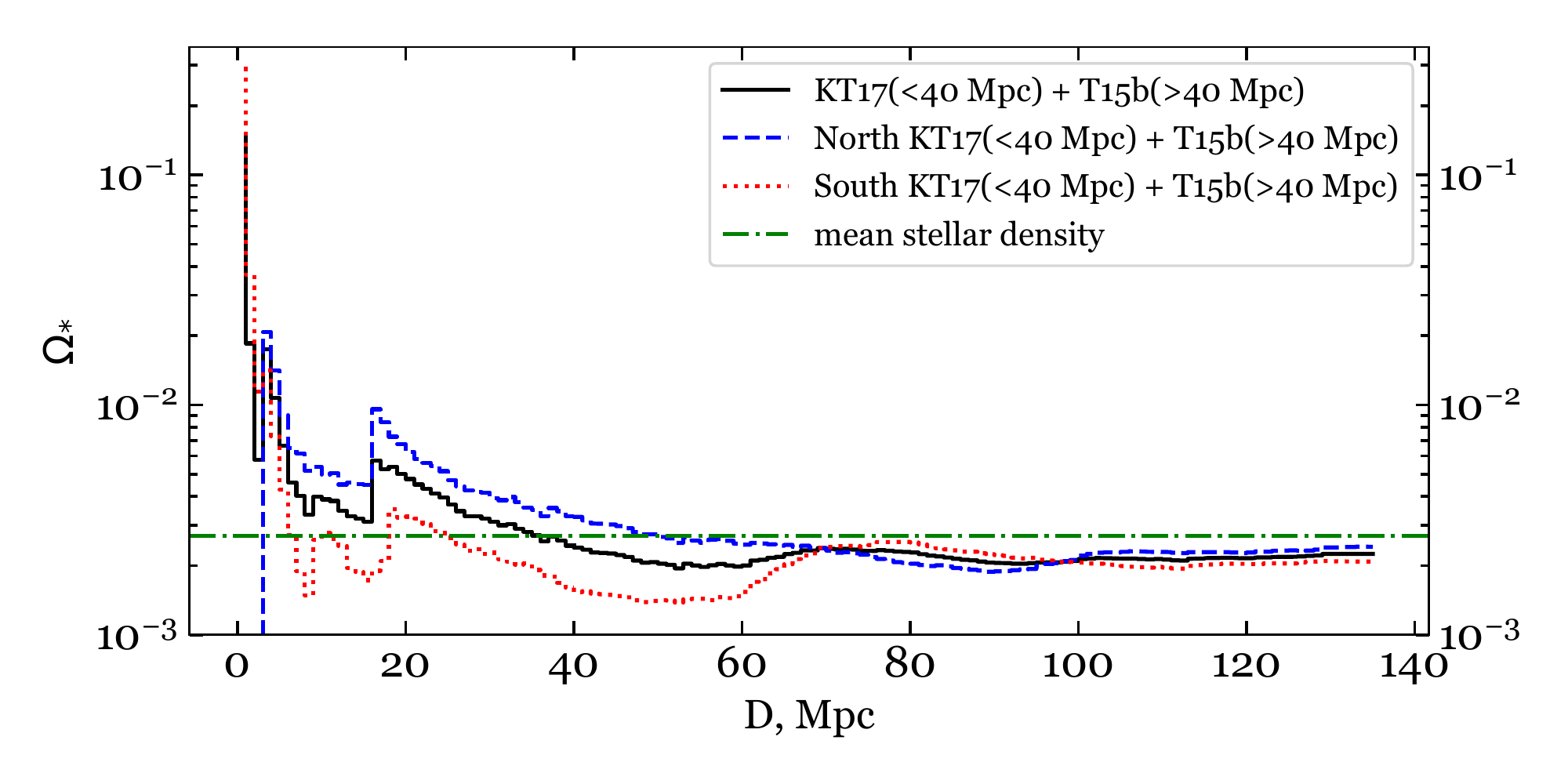}}
\caption{Mean stellar density calculated using KT17 catalog at $D<40$~Mpc and T15b catalog at $D=40-135$~Mpc. Middle curve corresponds to the stellar density for the whole sky, upper and lower curves match Northern and Southern hemispheres, respectively. The global stellar density is shown by dash-dotted horizontal line.} 
\label{7}
\end{figure*}
Figure~\ref{7} shows the mean stellar density profile calculated on the distance scales $D<135$~Mpc. To fill the nearby volume ($D<40$~Mpc) we used KT17 catalog. For farther distances we used T15b catalog with the correction factor CF. Notice, that values of $\Omega_*$ for Northern and Southern Galactic hemispheres are approximately equal to each other since $D>70$~Mpc. At the volume edge ($D=135$~Mpc) $\Omega_*$ is $(0.22\pm0.02)$\% of the critical density, being slightly lower than its global value ($0.27\pm0.05)$\% by \cite{Fukugita2004}. The difference between these quantities looks quite expected because 2MASS Survey misses about $20-25$\% of the total $K$-band luminosity.

\begin{figure*}[!ht]
\centerline{\includegraphics [width=342pt]{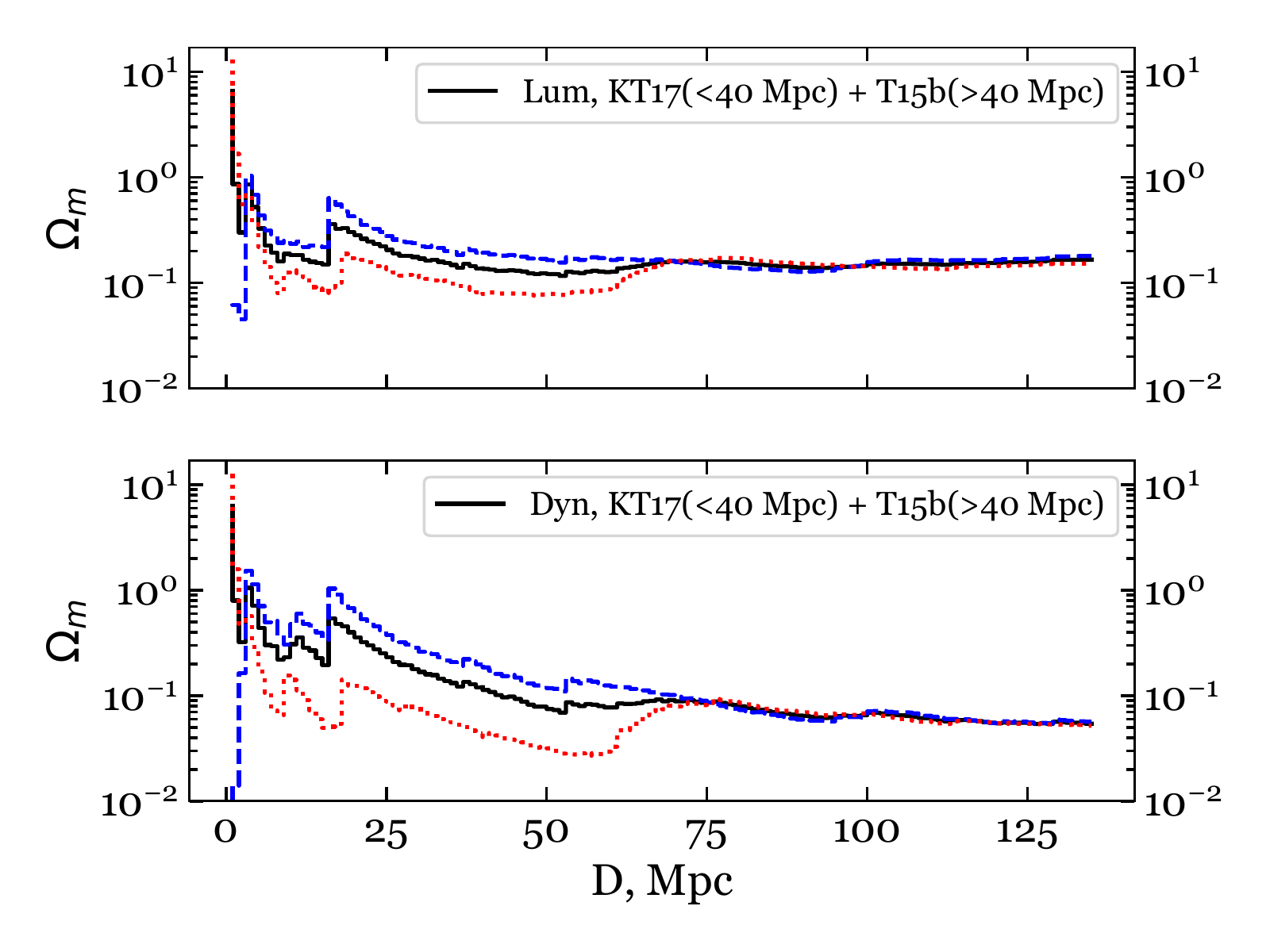}}
\caption{Mean density of the viral mass inferred from group's luminosity (upper panel) and from the dynamic mass (lower panel). Used catalogs are: KT17 at $D<40$~Mpc and T15b at $D=40-135$~Mpc. Middle curves correspond to the mean density for the whole sky, upper and lower curves match Northern and Southern hemispheres.} 
\label{8}
\end{figure*}
Distribution of the $\Omega_m(D)$ based on KT17 and T15b catalogs is shown in Figure~\ref{8}. Its upper and lower panels show the mean density of the total mass estimated by eq.~(\ref{M_L_Tully}) with accounting the correction factor CF and by kinematic characteristics of the galaxy groups, respectively. 
 
From Figure~\ref{8} one can draw the following conclusions: 
\renewcommand{\labelenumi}{\alph{enumi}.}
\begin{enumerate}
\item Difference between $\Omega_m$ for Northern and Southern Galactic hemispheres decreases with increasing $D$ and at $D>70$~Mpc it becomes within $(10-15)$\%  of the mean value.
\item Difference between mean densities of the total mass calculated by empirical relation~(\ref{M_L_Tully}) and by kinematics of galaxy groups increases with $D$ and reaches a factor $2-3$ at $D>70$~Mpc.
\item Within a sphere of 135~Mpc radius, the mean density of matter amounts to $\Omega_m=0.05\pm0.002$ via $M_{dyn}$ and $\Omega_m=0.16\pm0.01$ via $M_{Lum}$.
The latter value is in a good agreement with the quantity $\Omega_{collapsed}=0.16\pm0.02$ published by T15b.
\end{enumerate}

\begin{center}
\begin{table*}[!ht]
\centering
\caption{\label{tab:results} Mean densities and total masses of the stellar and dark matter within spheres with radii $D$.}
\begin{tabular}{lcclll} 
\hline
$D$&Sample&$M_*(<D)$&$\Omega_*(<D)$&$M_{tot}(<D)$&$\Omega_m(<D)$\\
Mpc& &$M_{\odot}$&$\%$&$M_{\odot}$& \\   
\hline
11&UNGC (LV)&$2.7 \times 10^{12}$&0.44&$1.0 \times 10^{14}$&0.17\\	     
11&MK11&$3.0 \times 10^{12}$&0.50&$1.1 \times 10^{14}$&0.18\\
11&KT17,lum&$2.3 \times 10^{12}$&$0.39 \pm 0.11$&$1.1 \times 10^{14}$&$0.18 \pm 0.05$\\   
&KT17, dyn&&&$1.9 \times 10^{14}$&$0.31 \pm 0.17$\\  
40&MK11&$9.2 \times 10^{13}$&0.32&$2.7 \times 10^{15}$&$0.09\pm0.03$\\
40& KT17, lum&$7.1 \times 10^{13}$&$0.24\pm0.08$&$4.0 \times 10^{15}$&$0.14\pm0.06$\\ 
&KT17, dyn&&&$3.5 \times 10^{15}$&$0.12\pm0.08$\\
135&T15b, lym&$2.5 \times 10^{15}$&$0.22\pm0.02$&$1.8 \times 10^{17}$&$0.16\pm0.01$\\ 
&T15b, dyn&&&$0.6 \times 10^{17}$&$0.05\pm0.002$\\
\hline

\end{tabular}
\end{table*}
\end{center} 
Table~\ref{tab:results} summarizes estimation of the stellar and dark matter masses and corresponding mean densities within the spheres with radii of 11, 40 and 135 Mpc based on different catalogs. Mass uncertainties correspond to the half-difference between mass estimations for Southern and Northern Galactic hemispheres.
		   
\section{Discussion and conclusions}
\begin{center}
\begin{table*}[t]
\centering
\caption{\label{tab:Mass_Vol_fractions_from_simulations} The mass and volume fractions (in \%) occupied by cosmic web environments from simulations by \cite{Nuza2014} and \cite{Cautun2014MNRAS}.}
\begin{tabular}{lcccc} 
\hline
&\multicolumn{2}{c}{Mass fraction}&\multicolumn{2}{c}{Volume fraction}\\
&Nuza&Cautun&Nuza&Cautun\\
\hline
Voids       &      22&   15&71&   76\\
Walls      &       22&   24&20&   18\\
Nodes     &        22&   11&1&   0.1\\
Filaments&         34&   50&8&    6\\    
\hline

\end{tabular}
\end{table*}
\end{center} 
\begin{center}
\begin{table*}[t]
\centering
\caption{\label{tab:grav_lens}Searching for galaxy clusters based on weak gravitational lensing.}
\begin{tabular}{lccccl} 
\hline
Telescope&      Seeing&   Area&    $N_{\rm peak}$&  Identified&   Reference\\
&arcsec&   sq.deg.&&           per cent\\
\hline
 MPG/ESO, 2.2m&    0.90&     19&      17&       65&      \cite{Schirmer2007AA}\\
 CFHT,    3.9m&    0.71&     72&      51&       59&      \cite{Shan2012} \\
 Subaru,  8.5m&    0.56&    160&      65&       60&      \cite{Miyazaki2018}\\     
\hline

\end{tabular}
\end{table*}
\end{center} 
\vspace{-41.5pt} For reliable measurements of the mean density of matter within a nearby Universe we need deep photometric and spectroscopic surveys covering a major part of the sky. Moreover, in the Local Volume ($\sim10$~Mpc) we need also individual estimates of galaxy distances because their radial velocities are often distorted by a large peculiar component. Recent progress in the galaxy surveys like the Sloan Digital Sky Survey \citep{Abazajian2009} and Pan-STARRS \citep{Chambers2016} gives astronomers hope to improve these measurements.
According to observational data from HST, the mean stellar density within the Local Volume is $1.5-1.8$ times greater than the global cosmic value. Consequently, we live inside the positive baryonic matter fluctuation. Alongside this, in the same volume the mean density of dark matter is $\Omega_m=0.17-0.18$, i.e. less than its global value.

Within the volume of $D<40$~Mpc including the Local Supercluster and neighboring clusters, the mean stellar density is approximately equal to the global stellar density. Wherein, three independent estimates of the virial masses give $\Omega_m = 0.09-0.14$, the value which is $2 - 3$ times lower than $\Omega_m$ in the standard $\Lambda$CDM model.
		
Outside a sphere of radius $\sim50$~Mpc, 2MASS photometric survey and 2MASS Redshift Survey miss a significant part of the galaxies. This fact makes the estimates of $\Omega_*$ and $\Omega_m$ less certain. Using the T15b catalog with the correction factor CF for missed galaxies leads to the mean stellar density within $D=135$~Mpc nearly the same, $0.8\pm0.2$, as the mean global density. However, our calculations of the mean density of dark matter via dynamic masses of the groups or via the empirical halo-mass-to-luminosity relation yield the $\Omega_m$ value in the range from 0.05 to 0.16. Note, that the latter
value, 0.16, which is inferred from ``luminous mass'', is more trustworthy than the first one.
			
Using 264 objects from Cosmicflows-3 \citep{Tully2016} with accurately measured distances and radial velocities, \cite{Shaya2017} applied Numerical Action Method to calculate 3D-trajectories for galaxies, groups and clusters within a distance of 40~Mpc. Authors conclude that their model with parameters  $\Omega_m=0.244$ (WMAP) and $H_0=75$ km~s$^{-1}$~Mpc$^{-1}$ is in a good agreement with the existing observational data if they add into their model a dispersed (orphan) dark matter component with $\Omega_{orphan}=0.077\pm0.019$, distributed outside the virial zones of groups and clusters.

\cite{Nuza2014} presented distribution of matter in the Local Universe based on N-body simulations accounting for $\Lambda$CDM model with $\Omega_m=0.27$. They estimated a cosmic variance on the scale of 80~Mpc to be $\sim3$\%. The same level corresponds to the concept of homogeneous cosmic cell, but on the other hand, conflicts with the existence of huge structures like Shapley Supercluster \citep{Munoz2008}. 
\cite{Nuza2014} derived the mass and volume fractions occupied
by different components of the cosmic web and concluded that the predominant part of the cosmic matter is associated with filaments (34\% of the total mass), while nodes (clusters), walls and voids account for 22\% of the total mass each. \cite{Cautun2014MNRAS} used the $\Lambda$CDM Millenium N-body simulations and obtained even 50\% mass fraction of dark matter in filaments (see Table~\ref{tab:Mass_Vol_fractions_from_simulations}).

Finally, \cite{Nuza2014} estimated the fraction of dark matter concentrated within the virial zones of  groups and clusters in the Local Universe ($D<40$~Mpc). Their value, $\Omega_m=0.08$, is well consistent with observational data by \cite{Karachentsev2012}.
   
There are at least three ideas how to explain the difference between the local density $\Omega_m\simeq0.08$ and the global quantity $\Omega_m=0.24-0.31$:

\begin{enumerate}

\item In contrast to luminous matter, the dark matter extends far beyond the virial radius of a group
or cluster, and only a half of their total mass resides within the virial radius \citep{Rines2006AJ,Masaki2012ApJ}.

\item Our Galaxy with the Local Supercluster locates inside a giant void.
\item Major mass of the dark matter is unconnected with groups and clusters and extends between
them as dark filaments or like uniform ``ocean''.
\end{enumerate}

Let us briefly outline recent observational data concerning to each of them.

\begin{enumerate}
\item In recent years, we have been able to measure the total mass of nearby groups and clusters not only by internal (virial) galaxy motions, but also by motions of surrounding galaxies, which are retarding by overdensity. In this case the estimate of the total mass corresponds to the zero-velocity radius $R_0$, which exceeds $3-4$ times the virial radius $R_{vir}$. Analysis of the Hubble flow around the Virgo cluster shows that the total mass of the cluster inside $R_0$ is almost the same as its virial mass \citep{KarachentsevTully2014,Kashibadze2018}. Similar result has been obtained by \cite{Kashibadze2018} for the Local Group and other nearby groups from analysis of their Hubble flows. Consequently, the assumption about the existence of dark massive halos around local groups and clusters extending to $\sim(3-4)R_{vir}$ is not confirmed by observations.

\item Another idea is an assumption that we live inside the giant void \citep{Shafieloo2009,Romano2012}. Some recent observations favour the existence of such extended zone ($\sim200$~Mpc) with the mean stellar density about $15-40$\% less than that of the global value \citep{Keenan2013,Whitbourn2014,Bohringer2015}. However, another observations in the $K$-band not prove a significant local underdensity \citep{Djorgovski1995,Bershady1998,Totani2001}. Anyway, one needs the presence of a deep large void of $\sim200$~Mpc diameter to explain the observed threefold difference between the global and local $\Omega_m$ quantities. But the existence of such an extended structure disagrees with the common concept of the large-scale homogeneity of the Universe.
			    
\item At present, the most promising explanation of the $\Omega_m$-paradox is the assumption that the considerable fraction $(\sim2/3)$ of the dark matter is dispersed outside the virial and collapsing zones of galaxy groups and clusters. Diffuse non-virialized structures, like cosmic filaments and walls, can manifest themselves via effects of weak gravitational lensing. Methodology of searching for dark massive attractors by the weak gravitational lensing effects has been applied so far only to rich galaxy clusters \citep{Shan2012,Miyazaki2018}, but not to dark walls or filaments. Apparently, the most easily observable can be the dark filaments oriented along the line-of-sight.  Table~\ref{tab:grav_lens}
presents some results of searching for galaxy clusters based on weak gravitational lensing analysis that have been performed with three telescopes under the excellent seeing. As it follows from the data,
about $40\%$ of the detected peaks remain to be not still identified  with optical counterparts.

 A future observational program of searching for possible massive dark attractors will need wide-sky surveys with large telescopes providing a sub-arcsecond seeing.
\end{enumerate}

Finally, one can presume that apart from usual dark matter component, concentrated in galaxy systems, there is another dark matter medium of unknown nature, uniformly filling the intergalactic space. However, this idea is not compatible with the established cosmological paradigm and hence will be rejected by ``Occam's razor''. 

We are grateful to the referee for useful comments.
\section*{Acknowledgments}

This work was supported by the \fundingAgency{Russian Foundation for Basic Research}, grant \fundingNumber{18-02-00005}.


\end{document}